\documentclass[journal,10pt,twocolumn]{IEEEtran}
\usepackage{times}
\usepackage{epsfig}
\usepackage{graphicx}
\usepackage{amssymb}
\usepackage{amsmath}
\usepackage{algorithm}
\usepackage{algorithmic}
\usepackage{multirow}
\usepackage{multicol}
\usepackage{subfigure}
\usepackage{mathptmx}       
\usepackage{helvet}         
\usepackage{courier}        
\usepackage{type1cm}        
\usepackage{makeidx}         
\usepackage[bottom]{footmisc}
\usepackage{array}
\usepackage{cite}
\newcolumntype{L}[1]{>{\raggedright\let\newline\\\arraybackslash\hspace{0pt}}m{#1}}

\usepackage{array}
\usepackage{hyperref}

\usepackage{color}

\newcommand{\remove}[1]{}

\begin{document}
{\title{Vehicular Communications: Survey and Challenges of Channel and Propagation Models
}}
\author{{\large Wantanee Viriyasitavat$^{\dagger\mathsection}$, Mate Boban$^{\ddagger}$, Hsin-Mu Tsai$^{*}$, and Athanasios V.~Vasilakos$^{\star}$}\\
\smallskip
\smallskip
\smallskip
{\small $^\dagger$Faculty of Information and Communication  Technology, Mahidol University, Thailand \\
\smallskip
$^\mathsection$Department of Telematics, Norwegian University of Science and Technology, Trondheim, Norway \\
\smallskip
$^\ddagger$NEC Laboratories Europe, NEC Europe Ltd., Kurf\"ursten-Anlage 36, 69115, Heidelberg, Germany\\
\smallskip
$^*$Department of Computer Science and Information Engineering, National Taiwan University, Taipei, Taiwan\\
\smallskip
$^{\star}$Department of Computer Science, Lulea University of Technology, Sweden\\
\smallskip
E-mail: wantanee.vir@mahidol.ac.th, mate.boban@neclab.eu, hsinmu@csie.ntu.edu.tw, vasilako@ath.forthnet.gr}}
\maketitle
\maketitle

\begin{abstract}
Vehicular communication is characterized by a dynamic environment, high mobility, and comparatively low antenna heights on the communicating entities (vehicles and roadside units). These characteristics make the vehicular propagation and channel modeling particularly challenging. In this survey paper, we classify and describe the most relevant vehicular propagation and channel models, with a particular focus on the usability of the models for the evaluation of protocols and applications. We first classify the models based on the propagation mechanisms they employ and their implementation approach. We also classify the models based on the channel properties they implement, where we pay special attention to the usability of the models, including the complexity of implementation, scalability, and the input requirements (e.g., geographical data input). We also discuss the less-explored aspects in the vehicular channel modeling, including modeling specific environments (e.g., tunnels, overpasses, parking lots) and types of communicating vehicles  (e.g., scooters, public transportation vehicles). We conclude the paper by identifying the under-researched aspects of the vehicular propagation and channel modeling that require further modeling and measurement studies.
\end{abstract}

\section{Introduction}	
\label{intro}

\begin{figure}[t]
\includegraphics[trim=10cm 22cm 10cm 22cm,clip=true,width=3in]{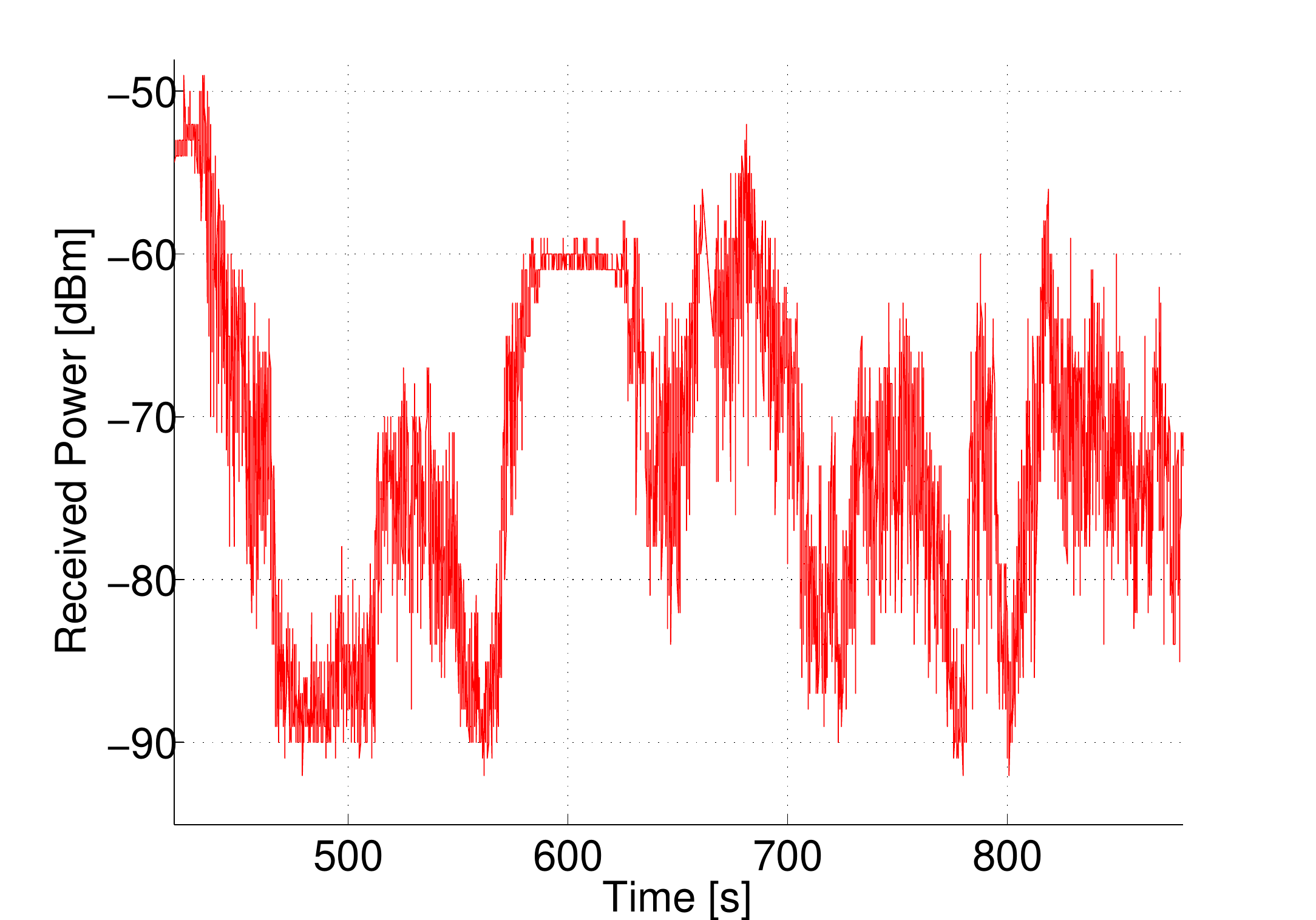}
\caption{Received power measurements at 5.9~GHz for V2V communication in an urban environment.}
\label{fig:measurementsPorto}       
\end{figure}

\begin{figure}
  \begin{center}
    \includegraphics[width=0.45\textwidth]{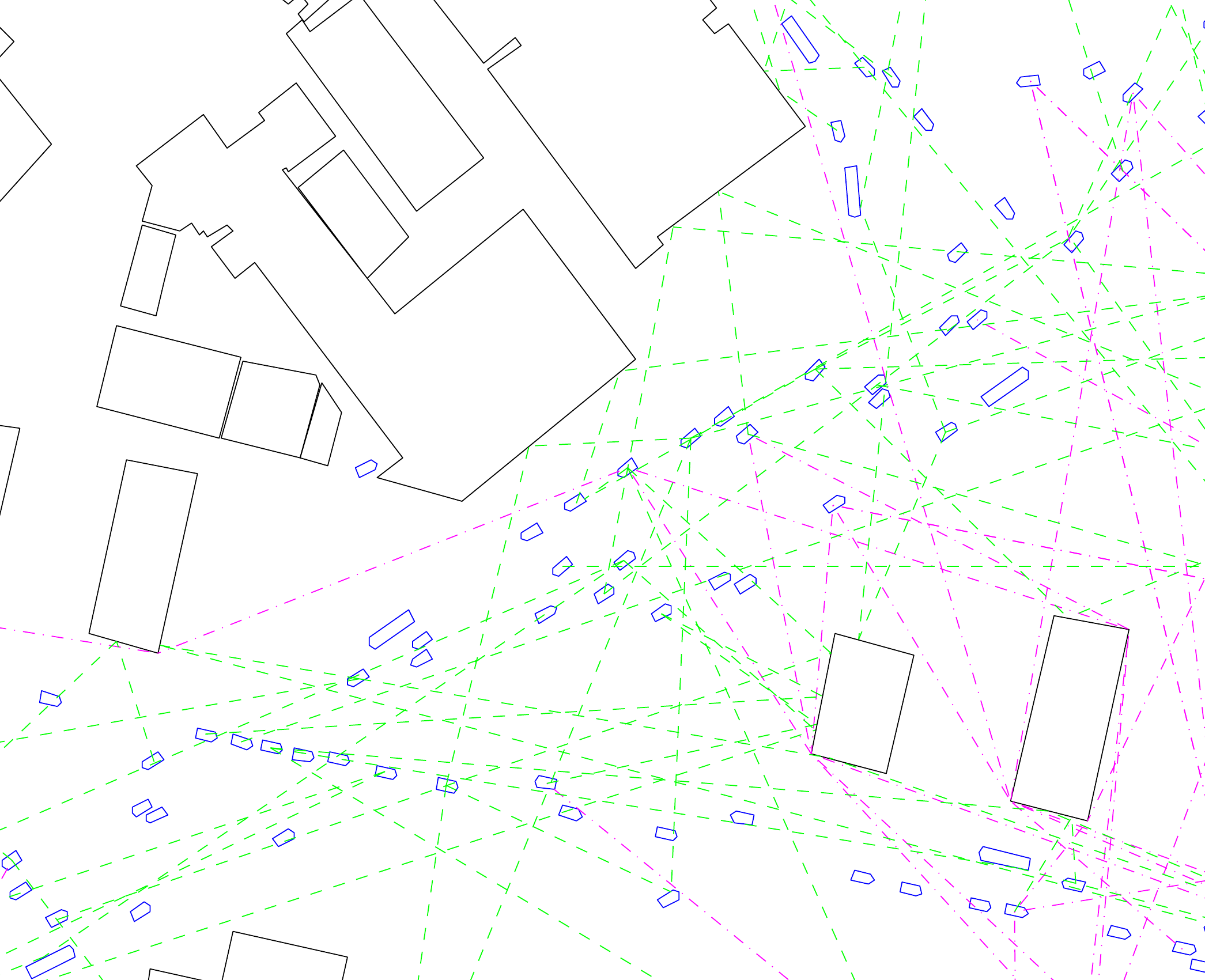}
     \caption{ 
     Simulation of propagation mechanisms in an urban area. Reflections and diffractions are shown for randomly selected communication pairs. Objects in the scene: buildings (black lines), vehicles (blue lines), reflected rays (green dashed lines), diffracted rays (magenta dash-dotted lines). 
} 
     
      \label{fig:PortoReflDiffr10kPairs}
   \end{center}
\end{figure}

The most important characteristics that separate vehicular communications, and therefore the vehicular channel modeling, from other types of wireless communications are: a) diverse environments where the communication happens; b) combinations of different communication types: vehicle-to-vehicle (V2V), vehicle-to-infrastructure (V2I), vehicle-to-pedestrian (V2P), etc.; and c) the objects, both static and mobile, that affect the vehicular communication. In combination, these characteristics result in complex propagation environments that are a challenge to model.  
Fig.~\ref{fig:measurementsPorto} shows how the small- and large-scale signal statistics vary rapidly in a typical urban environment, due to the dynamic environment, low height of the antennas, and high mobility of vehicles. Looking into the propagation characteristics, Fig.~\ref{fig:PortoReflDiffr10kPairs} shows that the built-up nature of the environment causes the signal traversing from the transmitter to the receiver to interact with a large number of surrounding objects. Even for single bounce (e.g., first-order) reflections and diffractions in urban environments, the number of resulting rays at the receiver is large. 
High density of objects, combined with the high mobility of the communicating vehicles and their surroundings, shows that capturing the characteristics of vehicular channels is far from trivial. 

While a number of existing mobile channel models have been extensively used for cellular systems, they are often not well suited for vehicular systems, due to the unique features of vehicular channels pointed out above. For instance, differences in the relative height of the transmitter and receiver antennas could lead to significantly different signal propagation behavior. The operating frequency and communication distance in vehicular communications also differ from those of cellular systems. Vehicular communication systems are envisioned to operate at 5.9 GHz and over short distances (10-500~m), whereas currently deployed cellular systems operate at 700-2100 MHz over a long distance (up to tens of kilometers)~\cite{molisch2009survey}.

There exist a number of surveys on V2V channel models. For example, Molisch et al.~\cite{molisch2009survey} describe key issues in V2V channels and summarize the V2V channel measurement studies in various scenarios. The authors classify V2V channels based on their implementation approach and discuss the advantages and disadvantages of each approach. Mecklenbrauker et al.~\cite{mecklenbrauker2011vehicular} review both the V2V and V2I propagation channels, with the main 
focus on the impact of different vehicular channel characteristics on the design of a vehicular wireless system. Cheng et al.~\cite{cheng09} survey V2V channel measurement and models, including 
the model classification based on the implementation approach. In addition, the authors also suggest guidelines for setting up a V2V measurement and developing realistic V2V channel models.

We survey the state of the art in vehicular channel modeling with a particular focus on: i) usability of the models for simulation at different scales (e.g., link-level vs. system-level) and taking into account the amount of geographic information available; ii) specific issues that need to be considered for the actual deployment of vehicular communication systems; and iii) providing guidelines in choosing a suitable channel model. 
Because the rollout of vehicular communication systems that is planned in the coming years in EU, U.S., Japan, and other countries\footnote{with the finalization of Release 1 standardization package by ETSI and CEN/ISO following the EC mandate M/453~\cite{euPress} and the recent announcement by the U.S. Department of Transportation to move forward with V2V communication~\cite{usdotNHTSA}},  large-scale evaluation and fine-tuning of standardized protocols and applications before their deployment has become the primary focus of simulation campaigns.  
For this reason, we pay particular attention to the usability aspects of vehicular channel models. In other words, we investigate whether or not the state of the art models can be used for efficient simulations of vehicular communication systems on a large scale. Using an appropriate channel model is critical for accurately evaluating vehicular protocols and applications before the actual deployment. To that end, we provide guidelines for choosing a suitable channel model, depending on the type of protocol/application under evaluation, available geographical information, and time constraints with respect to the simulation execution.

\begin{figure*}
  \begin{center}
\subfigure[ Urban area: high-rise buildings, moving vehicles, parked vehicles, occasional foliage.]{\label{fig:DowntownPortoRoute}\includegraphics[height=0.18\textwidth]{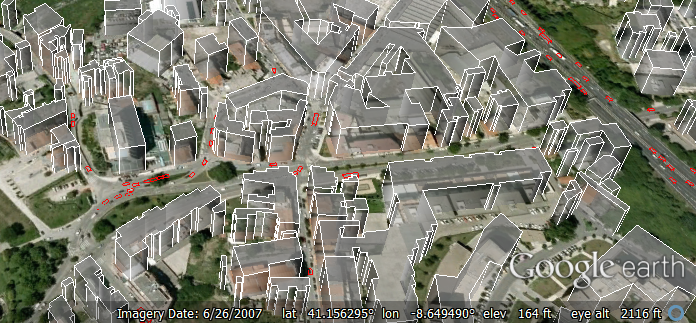}}\hspace{1mm}
\subfigure[ Suburban area: low-rise buildings, moving vehicles, frequent foliage.]{\label{fig:LecaRoute}\includegraphics[height=0.18\textwidth]{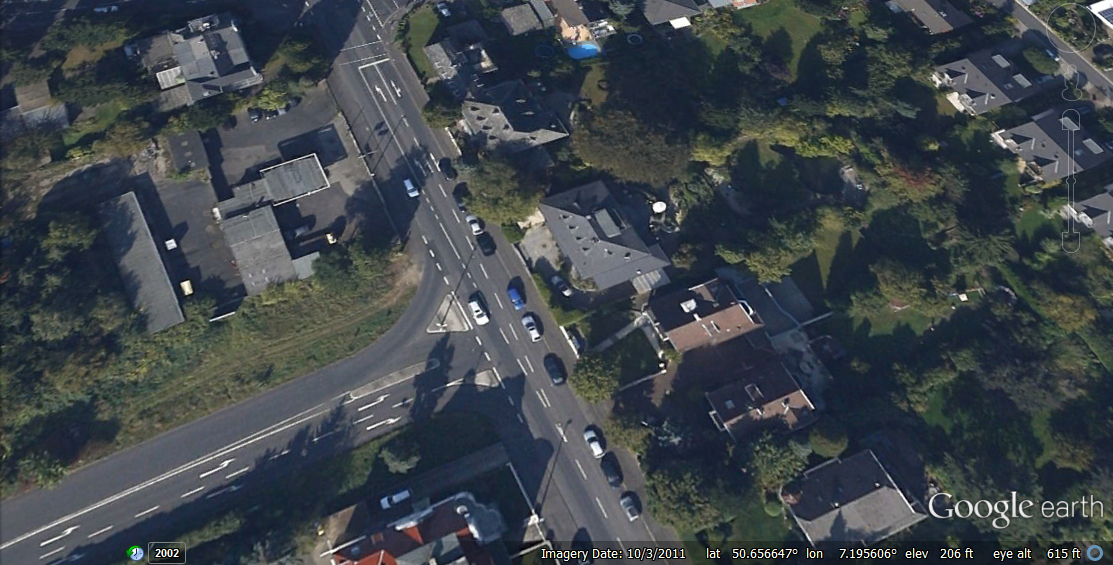}}\hspace{1mm}
\subfigure[ Highway. moving vehicles.]{\label{fig:VCIImgRoute}\includegraphics[height=0.18\textwidth]{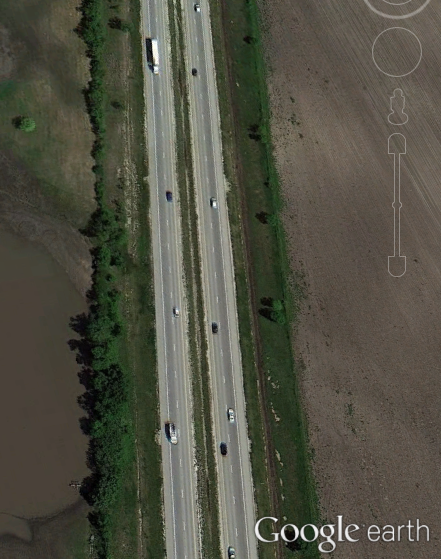}}  \hspace{1mm}
     \caption{Qualitative classification of typical vehicular communication environments and dedicated propagation obstacles.}%
      \label{fig:Environments}
   \end{center}
\end{figure*}

\section{Specific Considerations for Vehicular Channels}
\label{sec:Considerations}


\subsection{Environments}

The radio propagation is strongly influenced by the type of environment where the communication occurs. In case of vehicular communications, the most important objects that influence the propagation are buildings, vehicles (both static and mobile), and different types of vegetation. A combination of different object types, as well as their number, size, and density, has a profound impact on the radio propagation. While identifying different object types is not difficult, the classification of vehicular environments that they create is not a trivial task.  
Therefore, the environments where vehicular communication occurs are most often qualitatively classified as highways, suburban areas, and rural areas. Fig.~\ref{fig:Environments} shows the most often analyzed propagation environments. Varying presence, locations, and density of roadside objects as well as the velocity and density of vehicular traffic significantly impacts the signal propagation in these environments.  Therefore, the classification of environments should be taken with a grain of salt, because it is not uncommon to have an urban area that has open spaces akin to highways, or neighborhoods with low-rise buildings which could be arranged similar to a typical suburban setting. This is confirmed by numerous measurement studies, which have reported highly variable and often contradicting path loss exponents for the same environment: $1.6-2.9$ on highways~\cite{abbas12,karedal09,paschalidis11}, $2.3-3.5$ in suburban environment~\cite{Cheng2007,paschalidis11}, and $1.8-3.4$ in urban environment~\cite{mangel11_2,paschalidis11}. Similarly, mean delay spreads ranging between $140-400$~ns, $80-104$~ns, and $150-370$~ns have been reported for highways, suburban, and urban scenarios, respectively~\cite{renaudin2008wideband,acosta07,tan08}. Therefore, designing the propagation models with a specific environment in mind cannot ensure that the model will accurately apply to a different environment of the same ``class''. For this reason, the preferable method is to design propagation models that take into account the specific objects in the environment, along with their accurate dimensions and locations.


\subsection{Link Types} 
In addition to the nature of the propagation environment, it is also important to distinguish between different link types, as they exhibit vastly different propagation properties. In V2V channels, the transmitter and receiver antennas are usually mounted on the vehicle rooftop and both vehicles are mobile, 
whereas in V2I channels, the base station (or access point) is stationary and may be elevated. V2P communication links are envisioned to support Vulnerable Road User safety applications~\cite{anaya2014vehicle}. Differences in mobility, shadowing, and relative height of the transmitter and receiver antennas create significant differences in reflections, diffractions and scattering patterns of the transmitted waves~\cite{rappaport96}. 

\subsection{Vehicle Types}
Different types of vehicles (e.g., personal vehicles, commercial vans, trucks, scooters, and public transportation vehicles) have distinct dimensions and mobility dynamics. Therefore, models for the propagation characteristics of one vehicle type is not readily applicable to other types. Distinct features of vehicle types have an impact on the propagation modeling even if the vehicle itself is not the transmitter or the receiver. For example, the additional attenuation caused by a large truck blocking the line-of-sight (LOS) between the transmitter and the receiver can be more than 20~dB higher than the attenuation caused by personal vehicles~\cite{boban14TVT,vlastaras2014impact}.

\subsection{Objects}
Regardless of the link types, vehicular propagation environments also consist of a number of different types of objects that impact the signal propagation. The level of impact varies depending on the object type, the link type, and the environment. For instance, mobile objects (i.e., vehicles on the road) are more important for modeling vehicular channels in highway environments, because the communication between the transmitting and receiving vehicles on highways usually happens over the road surface. On the other hand, in urban environments with two-dimensional topology, the communicating vehicles are likely to be on different streets. In this case, along with mobile objects, accounting for static objects is critical for modeling vehicular channels, since both types of objects are sources of shadowing, reflections, and diffractions~\cite{maurer04_2}.

A number of measurement campaigns have also indicated that the LOS condition is a key factor in modeling V2V propagation channels. For example, measurements performed by Tan et al.~\cite{tan08} have shown that, regardless of the propagation environment (e.g., highway or urban scenarios), non-LOS channels have noticeably larger root-mean-square (RMS) delay spreads than that of LOS channels. This is due to the stronger signal attenuation and multipath effects caused by an increasing number of reflections and diffractions.

\section{Classification and Description of Vehicular Channel Models}
\label{sec:ModelClass}
In this section, we give an overview and recent advances in the vehicular propagation and channel modeling. The models described in this section are chosen mainly based on their usability (e.g., scalability, database input requirements, extensibility to different environments) and ability to realistically model a wide range of environments. By and large, the models presented in this section have been validated against measurements. We classify the models based on the propagation mechanism they model, the implementation approach they employ, and the channel properties they implement.

\subsection{Propagation Mechanisms}
Key distinguishing aspects of vehicular channels are varying path loss across space (e.g., different environments) and time (e.g., different time of day), potentially high Doppler shift, and frequency-selective fading caused by both mobile and static objects. Because modeling all of these aspects is a complex task, the most common approach thus far has been piecemeal modeling, wherein the problem is split into manageable parts and modeling is performed on one or more parts. 

\subsubsection{Large-scale propagation}
The most commonly used large-scale propagation model for vehicular channels is the {\it log-distance path loss} model~\cite{rappaport96}, with the associated 
path loss exponent being estimated based on empirical measurements. Cheng et al.~\cite{Cheng2007} fit the dual-slope log-distance model with suburban channel measurements. Similar approaches are used in various scenarios, including highways~\cite{abbas12}, rural and highway scenarios~\cite{karedal09}, urban intersection scenarios~\cite{mangel11_2}, and garage scenarios~\cite{sun2013parking}. In addition, other large-scale models are used. Geometry-based Efficient propagation Model for V2V communication (GEMV$^2$) proposed by Boban et al.~\cite{boban14TVT} uses different types of path loss models for LOS and non-LOS conditions (the two-ray ground reflection model~\cite{rappaport96} and log-distance path loss, respectively), whereas the model proposed by Maurer et al.~\cite{maurer04_2} uses the ray-tracing techniques~\cite{00parsons} to model the large-scale propagation effects.

\subsubsection{Small-scale fading}
In addition to large-scale propagations, a number of models have been proposed to account for the small-scale signal variations caused by multipath propagations and Doppler effects due to  mobility of vehicles and objects in their surroundings. Similar to the large-scale propagation modeling, the small-scale fading is usually modeled using well-known distributions such as the Weibull~\cite{sen08}, Nakagami~\cite{Cheng2007}, and Gaussian~\cite{boban14TVT} distributions with parameters estimated from the measurement data. For instance, in the GEMV$^2$ model~\cite{boban14TVT}, the small-scale fading is modeled using the Gaussian distribution with varying standard deviation depending on the number of vehicles and density of objects in the area. Ray-tracing techniques have also been used to estimate the small-scale fading in various environments~\cite{karedal09,maurer04_2}.

It is worth noting that the propagation characteristics of vehicular communications are highly dependent on 
the existence of the LOS path, as indicated by empirical measurements (e.g.,~\cite{boban14TVT,abbas12}). As a result, the large- and small-scale propagation characteristics are usually modeled separately for LOS and non-LOS links. Mechanisms to differentiate the link types (e.g., LOS, non-LOS due to vehicles, non-LOS due to buildings) are included in recent models~\cite{boban14TVT,sun2013parking,abbas12}. 

\subsection{Channel Model Implementation Approaches}
Depending on the implementation approach and the availability of geographical information, the models can be classified based on their implementation approach as follows.

\subsubsection{Geometry-based (GB) models}
\begin{itemize}
	\item \textbf{Ray-tracing models}~\cite{00parsons} are the most commonly-used geometry-based deterministic (GBD) models for the vehicular channel modeling. Ray-tracing methods require a detailed description of the propagation environment to produce the actual physical propagation process for a given environment in order to accurately calculate the channel statistics. The model proposed by Maurer et al.~\cite{maurer04_2} is an example of a model that is based on ray-tracing. It calculates the channel statistics by analyzing the 50 strongest propagation paths between the transmitter and the receiver. A more scalable RAy-tracing Data Interpolation
and Interfacing model (RADII) is proposed by Pilosu et al.~\cite{pilosu2011radii}. RADII uses a combination of pre-processing ray-tracing techniques to compute the average attenuation of each region of interest (ROI) and uses an interpolation technique to compute attenuation between connected ROIs offline, so that the simulations can use a lookup table without the need of recalculating the channel statistics.
	\item \textbf{Simplified geometry-based models} take into account the geometric properties of the surroundings, at the same time simplifying geometric calculations by
	extracting 
	some of the channel statistics 
	either from measurements or simulations. Examples of these models have been described by 
	Cheng et al.~\cite{Cheng2007} and Sun et al.~\cite{sun2013parking}. In these models, channel parameters are estimated separately for a given measurement scenario. For instance, two sets of model parameters are estimated for two suburban environments in Cheng et al.~\cite{Cheng2007} and several sets of parameters are estimated in a parking garage environment by Sun et al.~\cite{sun2013parking}, 
	depending on the LOS/non-LOS condition and the locations of the transmitter and the receiver. Karedal et al.~\cite{karedal09} propose a more complex model that takes into account four distinct signal components: LOS, discrete components from mobile objects, discrete components from static objects, and diffuse scattering. Model parameters were extracted from measurements in highway and suburban environments. Abbas et al.~\cite{abbas12} designed a model that differentiates LOS and non-LOS conditions of a link based on a Markov chain probabilistic model. The transition probabilities between conditions are estimated from the probability distributions of the LOS and non-LOS components measured in different environments. Mangel et al.~\cite{mangel11_2} developed a channel model that incorporates relevant information about street intersections (e.g., street width, existence of buildings on intersection corners, etc.). The model is fitted to 
the measurements that the authors performed at representative intersections. Boban et al. \cite{boban14TVT} developed the GEMV$^2$ model, which 
uses outlines of vehicles, buildings, and foliage to distinguish three types of links: LOS, non-LOS due to other vehicles~\cite{boban11}, and non-LOS due to buildings or foliage.

	\end{itemize}
\subsubsection{Non-geometry-based (NG) models}

	Most NG models conform to the following recipe: measuring the channel characteristics in a specific environment and adjusting the parameters of the path loss, shadowing, and the small-scale fading accordingly. One of the most widely-used NG models is the tapped-delay line (TDL) model. Each tap in this model represents signals received from several propagation paths; each with a different delay and different type of Doppler spectrum. Based on an extensive measurement campaign performed in urban, suburban, and highway environments with two levels of traffic density (high and low), Sen and Matolak~\cite{sen08} proposed a TDL model for each region. The Markov chain technique is used to model the multipath component whereby the non-stationarity property of the model is incorporated by adding the {\it persistence} process which accounts for the finite ``lifetime'' of the propagation paths. Similarly, Wang et al. proposed a TDL-based channel model with birth/death processes to account for a sudden appearance of a LOS component~\cite{wang12}. 	
	
\subsection{Properties of the Model}

Since the focus of this paper is on the usability of the model for the protocol and application evaluations, we identify below the most important properties that enable the usability of the model. Based on these properties, Table~\ref{tab:Models} qualitatively summarizes the state of the art propagation and channel models. 

\begin{table*}[ht!]
\caption{Classification of Propagation and Channel Models. }
\label{tab:Models}      
\begin{tabular}{p{2cm}p{2.1cm}p{2cm}p{1.2cm}p{0.9cm}p{1cm}p{1cm}p{1cm}p{1.cm}p{0.8cm}p{1.5cm}}
\hline
\multirow{3}{*}{Model} & \multicolumn{2}{c}{Propagation scale} & \multirow{3}{*}{Environment}  & \multirow{2}{*}{Implem.} & \multicolumn{5}{c}{Properties of the channel models}  \\
\cline{2-3}\cline{6-11}
& \multirow{3}{*}{Large} & \multirow{3}{*}{Small} && \multirow{2}{*}{approach} & Spatial- & Non- & Extensi- & Applica- & Antenna & {Scalability} \\
&&&&&temporal&stationarity&bility&bility&config.&\& link\\
&&&&&depend.&&&&&complexity\\
 [1ex]
\hline
\multirow{2}{*}{Sun et al.~\cite{sun2013parking}} & \multirow{2}{*}{Log Distance}&\multirow{2}{*}{-} & Parking & \multirow{2}{*}{GBD} &  \multirow{2}{*}{-} & \multirow{2}{*}{-} & \multirow{2}{*}{-} & \multirow{2}{*}{-} & \multirow{2}{*}{-} & \multirow{2}{*}{Large, $O(1)$} \\
&& &garage & &&&&&&\\[2ex]

\multirow{2}{*}{Fayziyev et al.~\cite{Fayziyev2014channel}} & Measurement-&\multirow{2}{*}{-} & \multirow{2}{*}{Tunnel} & \multirow{2}{*}{GBD} &  \multirow{2}{*}{-}&  \multirow{2}{*}{-} & \multirow{2}{*}{-} & \multirow{2}{*}{-} & \multirow{2}{*}{-} & \multirow{2}{*}{Large, $O(1)$} \\
& fitted impulse res.& & & &&&&&\\[2ex]

\multirow{1}{*}{Abbas et al.~\cite{abbas12}} & \multirow{1}{*}{Log Distance}& \multirow{1}{*}{-*} & \multirow{1}{*}{Highway} & \multirow{1}{*}{GBS} & \multirow{1}{*}{-} & \multirow{1}{*}{-} & \multirow{1}{*}{-} & \multirow{1}{*}{-} & \multirow{1}{*}{-} & Large,  $O(1)$\\[2ex]

Sen and &  - & \multirow{2}{*}{Weibull} & Urban, & \multirow{2}{*}{NG} & \multirow{2}{*}{X}& \multirow{2}{*}{X} & \multirow{2}{*}{-} & \multirow{2}{*}{-} & \multirow{2}{*}{-} & \multirow{2}{*}{Large, $O(1)$}\\
Matolak~\cite{sen08} && &highway & &&&&&\\[2ex]

Wang et al.~\cite{wang12} & - & Rician & All & GBS & X &  X & X & X & - & Large, $O(1)$ \\[2ex]

\multirow{2}{*}{Cheng et al.~\cite{Cheng2007}} & Dual-Slope &\multirow{2}{*}{Nakagami}& \multirow{2}{*}{Suburban} & \multirow{2}{*}{NG} & \multirow{2}{*}{-}& \multirow{2}{*}{-} & \multirow{2}{*}{-} & \multirow{2}{*}{-} &\multirow{2}{*}{-} & \multirow{2}{*}{Large, $O(1)$}\\
&Log Distance&&&&&&&&&\\[2ex]

Mangel et al.~\cite{mangel11_2} & \multirow{2}{*}{Log Distance}& Nakagami (LOS)& \multirow{2}{*}{Intersections} & \multirow{2}{*}{GBD}& \multirow{2}{*}{X}& \multirow{2}{*}{-}& \multirow{2}{*}{X}& \multirow{2}{*}{-}&\multirow{2}{*}{-}&\multirow{2}{*}{Large, $O(1)$}\\
&&Normal (NLOS) & &&&&&&\\[2ex]

\multirow{2}{*}{Karedal et al.~\cite{karedal09}} &\multirow{2}{*}{Log Distance}& Simplified \;\;Ray-Tracing  &Rural, highway & \multirow{2}{*}{GBS} &\multirow{2}{*}{X}&  \multirow{2}{*}{X} & \multirow{2}{*}{X} & \multirow{2}{*}{-} & \multirow{2}{*}{X} & Medium, $O(R+V)$\\
&&&&&&&&&&\\[2ex]

\multirow{2}{*}{Boban et al.~\cite{boban14TVT}} & Two-Ray (LOS) & \multirow{2}{*}{Normal$^+$} & \multirow{2}{*}{All} & \multirow{2}{*}{GBD} &\multirow{2}{*}{X}&  \multirow{2}{*}{X}& \multirow{2}{*}{X} &\multirow{2}{*}{X}& \multirow{2}{*}{-}& \multirow{2}{*}{Large, $O(V)$} \\ 
&Log Dist. (NLOS)&&&&&&&&\\[2ex]

\multirow{2}{*}{Pilosu et al.~\cite{pilosu2011radii}} & Preprocessed& Preprocessed & \multirow{2}{*}{All} & \multirow{2}{*}{GBD} & \multirow{2}{*}{X}&\multirow{2}{*}{X} & \multirow{2}{*}{X} & \multirow{2}{*}{X} & \multirow{2}{*}{X} & Small\\
&Ray Tracing&Ray Tracing&&&&&&&&\scriptsize{$>O((R+V)^2)$}\\[2ex]

\multirow{2}{*}{Maurer et al.~\cite{maurer04_2}} & \multirow{2}{*}{Ray Tracing}& \multirow{2}{*}{Ray Tracing} & \multirow{2}{*}{All} & \multirow{2}{*}{GBD} &\multirow{2}{*}{X}& \multirow{2}{*}{X} & \multirow{2}{*}{X} & \multirow{2}{*}{X} & \multirow{2}{*}{X} & Small,\\
&&&&&&&&&&\scriptsize{$>O((R+V)^2)$}\\[2ex]
\hline
\vspace{0.01cm}
\end{tabular}
\textbf{NG}: Non-Geometry-based model, \textbf{GBS}: Geometry-Based  Stochastic model, \textbf{GBD}: Geometry-Based Deterministic model\\
$R$ and $V$ denote the number of roadside objects and vehicles, respectively. \\
*Only spatial correlation of shadow fading is considered. \\
$^+$Signal deviation depends on the number of vehicles and static objects in the area. 
\end{table*}

\begin{itemize}
\item \textbf{Spatial and temporal dependency}

While the small-scale fading models account for the time-varying signal attenuation due to propagation effects (e.g., reflections and scattering), measurements have demonstrated that the variation in signal attenuation is strongly correlated over both time and space. This spatial and temporal dependency arises from the static and dynamic physical world features, respectively. In other words, different communication links in an area are affected by the same effects (generated by, for example, obstructing objects and ambient noise/interference). These links exhibit similar characteristics due to spatial correlation. On the other hand, mobility of vehicles and varying traffic density lead to the signal attenuation that is correlated over time (i.e., temporal dependency). The ability of a channel model to include the spatial-temporal dependency is shown in Table~\ref{tab:Models}.\\

	\item \textbf{Temporal variance and non-stationarity}
	
In addition to the spatial and temporal correlation, measurements have revealed that vehicular channels exhibit the strong non-stationarity; i.e., in addition to a change in the channel state, the channel statistics may also change, especially if the channel involves vehicles that travel at high speeds. The non-stationarity of the model also arises from static and mobile objects that could cause a sudden appearance/disappearance of the LOS component. Table~\ref{tab:Models} classifies the channel models based on their ability to simulate the non-stationarity property of vehicular channels\footnote{Note that all of the models in Table~\ref{tab:Models} can simulate the temporal variance property of the channel.}.
		
	\item {\bf Extensibility to different environments}

	With regards to the applicability of a model to different environments, we distinguish between the channel models that were calibrated by extracting the pertinent parameters from measurements at a specific set of locations and those that have the ability to model effects beyond those captured at particular locations. Since the former category depends on measurements, 
	these models can give no accuracy guarantees for locations with considerably different characteristics. On the other hand, models that take into account geometry-specific information of the simulated area can give some insights for environments beyond those characterized by  measurements. 
For this reason, we indicate the extensibility of the model in Table~\ref{tab:Models} to describe whether or not the model can be generalized to other propagation environments beyond those that were used to generate the model.

	\item \textbf{Applicability}
	
	Since the primary purpose of vehicular channel models is to support the realistic development of vehicular and ITS-related applications, we analyze the ability of models to take into account application-specific scenarios. For example, instead of analyzing general highway scenarios, Bernad{\'o} et al.~\cite{bernardo13} performed measurements and subsequently developed channel models for different applications on highways: merging lane scenarios, traffic congestion scenarios, scenarios in which a car approaches a traffic jam, etc. While the classification by propagation environments can be used to identify some practical applications, there exist certain applications that require the dedicated channel characterization (e.g., pre-crash and post-crash warning~\cite{bernardo13}). In Table~\ref{tab:Models}, we identify the channel models that can be applied to other use-cases in addition to the ones they are originally calibrated for. 
	
	\item {\bf Antenna configuration }
	
	Related to channel model's ability to incorporate small-scale fading is the ability to support different types of antenna configurations that exploit the positive and counter the negative effects of small-scale fading. Therefore, we include the information about the model's ability to support different antenna configurations (e.g., Multiple-Input Multiple-Output antenna configuration). 

	\item \textbf{Scalability}
	
	In addition to the properties of the channel itself, we also classify the models based on their efficiency, which in turn determines the model's scalability. Given the increase in demand for efficient evaluations of vehicular applications, it is necessary for channel models to be able to support large-scale simulations. Efficiency of the model depends largely on the complexity of the mechanisms employed for calculating the channel statistics. In general, models that utilize the ray-tracing techniques can provide good accuracy but do not scale well. The scalability properties of the channel models are assessed qualitatively and shown in Table~\ref{tab:Models}. 
\end{itemize}

\subsection{Comparison of Selected Channel Models}
\begin{figure}[t]
\includegraphics[trim=5cm 21cm 6cm 22cm,clip=true,width=0.48\textwidth]{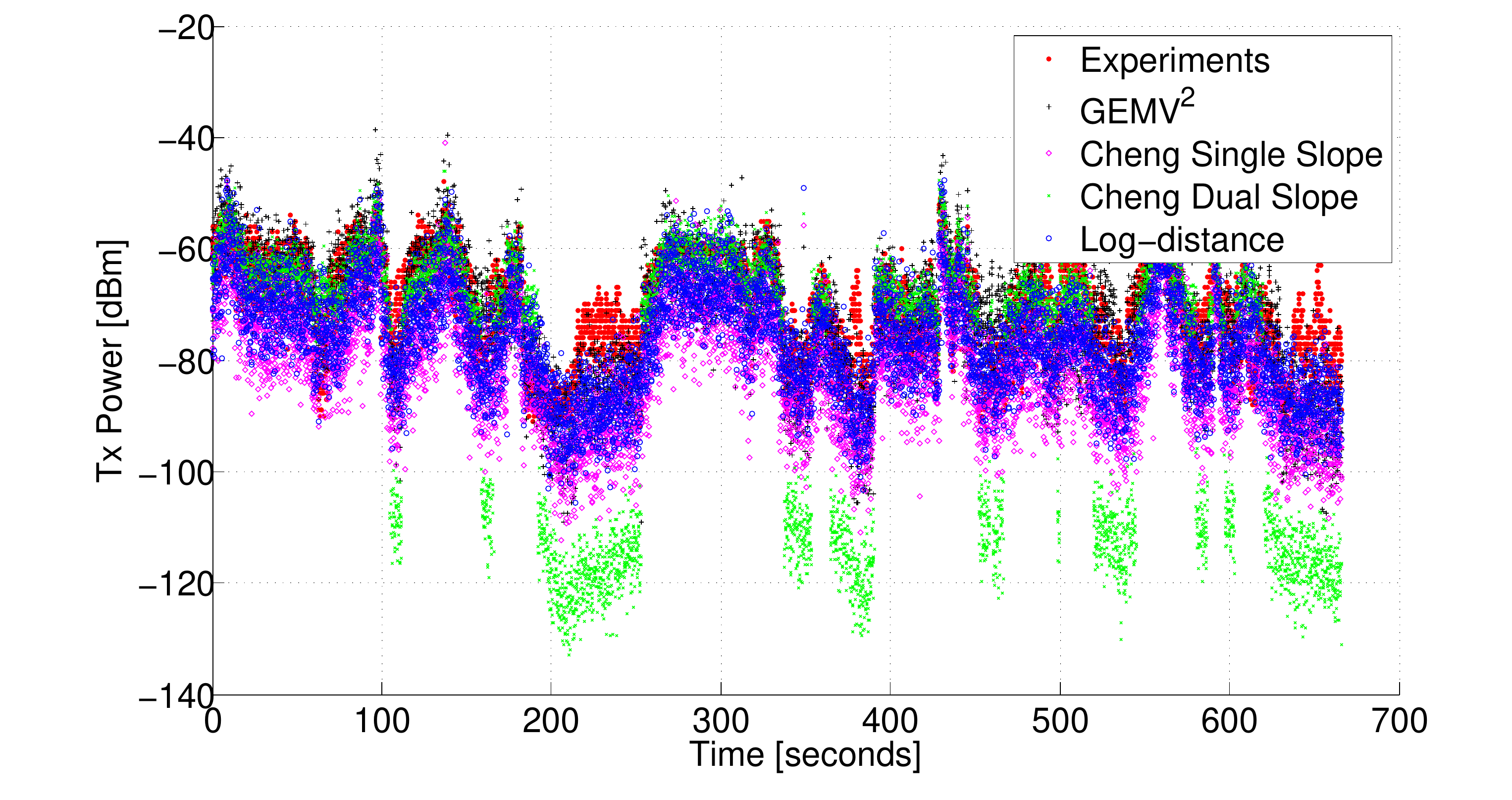}
\caption{Comparison of the received power results estimated by four models against the results obtained from V2V measurements performed in the city of Porto. Mean absolute error of each model (i.e., absolute difference for each measured data point): 6.7~dB (for the GEMV$^2$ model), 11.1~dB (for the Cheng single slope model), 14.4~dB (for the Cheng dual slope model), and 7.7~dB (for the log-distance model).}
\label{fig:measurementsPortoComparison}
\end{figure}

Figure~\ref{fig:measurementsPortoComparison} shows the comparison of the received power results obtained for a V2V measurement campaign performed in the city of Porto with four models: the GEMV$^2$ model~\cite{boban14TVT}, two models proposed by Cheng et al.\footnote{For the single slope model, path loss exponent of 2.75 and std. dev. for fading of 5.5 are used. For the dual slope model, path loss exponent of 2.1 is used for distance below 100~m and 3.8 for distance above 100~m; fading std. dev. of 2.6~dB is used for distance below 100~m and 4.4~dB above.}~\cite{Cheng2007}, and the log-distance path loss model with log-normal shadow fading\footnote{For the log-distance path loss model, path loss exponent of 2.5 and fading deviation of 5 dB are used.}. 
The parameters for the log-distance path loss model have been set to approximate the values extracted from the measurement data. Note that the actual locations of vehicles surrounding the communicating vehicles during the measurements are unknown. In case of the GEMV$^2$ model, this implies that their locations cannot be used in the model itself; instead, simulated locations were used, thus reducing the estimation accuracy of non-LOS links. 

The main take-away from the comparison is that, if the measurements for a specific environment are not available, then the NG models provide inconsistent results. For example, the path loss exponent for distance above 100~m in the dual-slope Cheng model is clearly too high, thus resulting in unrealistically low received power values above 100~m. If the model's parameters are extracted from the measurement data for a given location, then the estimate is better as shown by the log-distance model in Fig.~\ref{fig:measurementsPortoComparison}). However, if the geographical information is available, then the GBD models such as the GEMV$^2$ model are a better choice. 


\subsection{Guidelines for Choosing a Suitable Channel Model}
The models listed in Table~\ref{tab:Models} differ in many aspects and offer different trade-offs between accuracy and complexity/scalability. Stochastic models that do not require any information about the environment are simple and highly scalable, at the expense of lower accuracy. GB models trade off scalability for accuracy, where the trade-off can differ quite significantly from one model to another. 
Ray-tracing models (e.g.,~\cite{maurer04_2}) require a detailed information about the propagation environment (that can be hard to collect) and higher computational power. On the other hand, the model proposed by Abbas et al.~\cite{abbas12} only requires information on the type of the environment to estimate the channel statistics. Thus, it is highly scalable and can provide environment-specific {\it but not} location-agnostic channel information. Simplified GB models that take into account actual locations of objects (e.g., GEMV$^2$ model~\cite{boban14TVT}) can achieve a good accuracy/scalability tradeoff, offering a large gain on scalability compared to the ray-tracing models, while providing sufficient accuracy and ease of use. 

\begin{figure}[t]
\includegraphics[width=\columnwidth,height=2.75in]{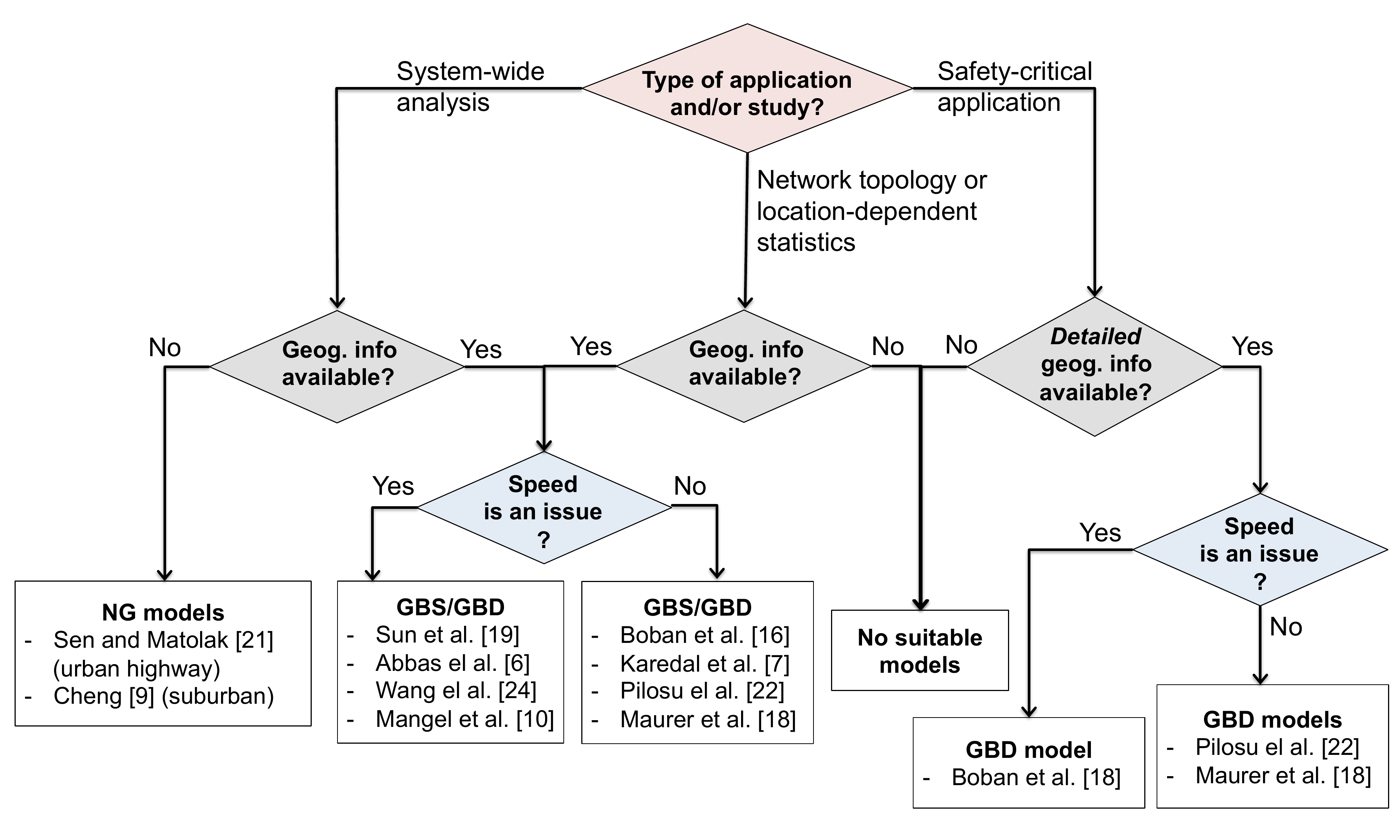}
\caption{Guidelines for choosing a suitable channel model.}
\label{fig:choosemodel}       
\end{figure}

Ultimately, choosing the right model should depend on the type of application and/or protocol that needs to be evaluated, constrained by processing power and availability of the required data (either geographical or measurements). 
To that end, the flowchart shown in Fig.~\ref{fig:choosemodel} provides a guideline in choosing a suitable channel model.
For example, if only system-wide performance analysis is required (e.g., overall packet delivery ratio, average end-to-end delay, etc.), any type of model (NG, GBS, or GBD models) might be suitable. However, if an application requires network topology statistics (e.g., the number of neighboring vehicles) or location-dependent statistics (e.g., the packet delivery rate or end-to-end delay in an area with rapid channel fluctuations), GB models that can model dynamic link transitions and small-scale variations should be used. For safety-critical applications, that disseminate time-sensitive information about a specific safety event, GBD models are the best choice.

Once the channel model category 
for a specific application is identified, the suitable channel model should be chosen based on the availability of geographic/measurement data and processing power. If the complete geographic information (e.g., location, dimensions, and material properties of vehicles, buildings, and foliage) is available and processing speed is not an issue, then ray-tracing-based GBD models (e.g., the model proposed by Maurer et al.~\cite{maurer04_2}) could be used for maximum accuracy. If limited information about the propagation environment is available (e.g., density of vehicles and surrounding objects) and the processing speed is important, then simplified GB models can be used (e.g.,~\cite{boban14TVT,karedal09}). Otherwise, other GB models such as~\cite{Cheng2007,wang12}, which only require a qualitative type of simulated environment may be used.

%


\section{Toward Realistic and Efficient Vehicular Channel Modeling}
In this section, we discuss the recent trends in the vehicular channel modeling, including the the need for models that are usable in large-scale vehicular network simulators. 
We also discuss vehicular channel emulation as an alternative approach for realistic protocol and application evaluations. Finally, we point out open problems in the area of propagation and channel modeling that require further attention.

\subsection{Efficient Models for Realistic Large-Scale Simulation}
As the deployment phase in main ITS markets is getting closer, realistic channel models for large-scale simulations are necessary for the efficient evaluation of applications before they are deployed in the real world. However, channel and propagation models currently used to simulate V2V and V2I communication links in VANET simulators (e.g., NS-3\footnote{\url{www.nsnam.org}}) are based on simple statistical models (e.g., free space, log-distance path loss~\cite{rappaport96}, etc.) that are used indiscriminately for all environments where the communication occurs. As shown in Fig.~\ref{fig:measurementsPortoComparison}, these models cannot capture the characteristics of vehicular channels, namely rapid transitions between LOS and non-LOS conditions, changes in delay and Doppler spreads, etc. Consequently, simple models were shown to exhibit poor performance in terms of link-level modeling, particularly in complex environments~\cite{dhoutaut06}. A way forward in this respect would be to combine geometry-based scalable propagation models (e.g.,~\cite{boban14TVT,abbas12}), which are able to distinguish between different LOS conditions and environments, with small-scale channel models, which are able to provide appropriate delay and Doppler statistics for each representative environment (e.g.,~\cite{Cheng2007,bernardo13}). Finally, attempts should be made to implement such realistic models in large-scale network simulators in order to enable realistic evaluations of protocols and applications. 
 
\begin{figure*}[t]
\centering
\includegraphics[width=0.95\textwidth]{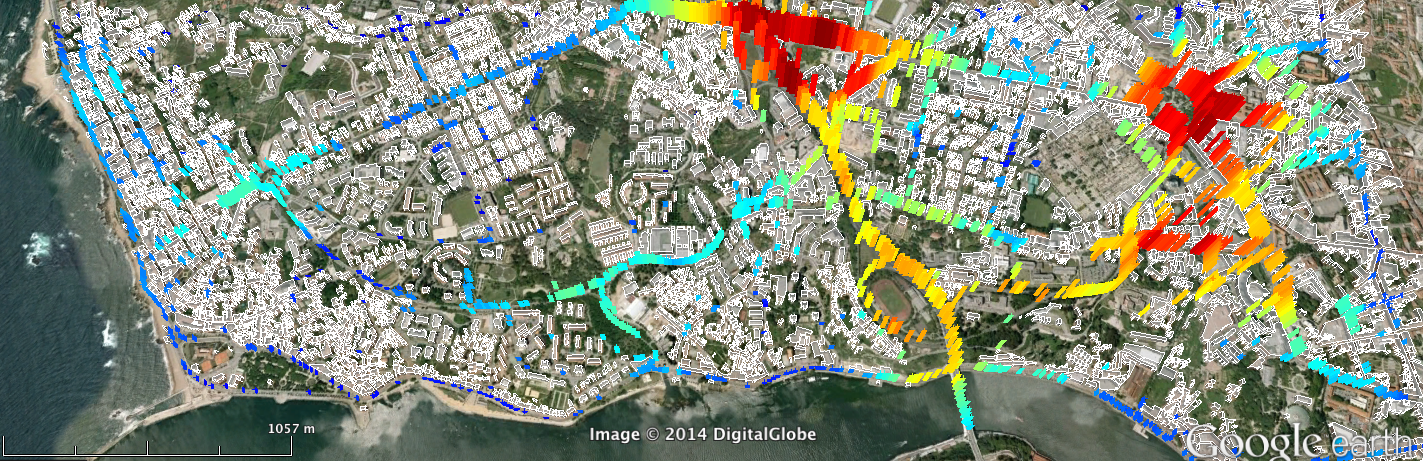}
\caption{Visualization of a neighborhood size generated by the GEMV$^2$ model. For each vehicle, the colored bar represents the number of vehicles it can directly communicate with (i.e., its neighbors). Warmer and taller bar colors indicate more neighbors.}
\label{fig:neighborhood}
\end{figure*}

The GEMV$^2$ model is an example of a computationally efficient channel model that can model the signal propagation in a large set of environments (e.g., highway, rural, urban, complex intersections, etc.) and is able to simulate city-wide vehicular networks with thousands of communicating vehicles. 
It allows importing realistic mobility data from Simulation of Urban MObility (SUMO) and building/foliage outlines from OpenStreepMap\cite{openstreetmap}\footnote{The source code of the GEMV$^2$ model is available at \url{http://vehicle2x.net/}.}. Apart from the propagation-related statistics, the GEMV$^2$ model allows for the analysis of networking related metrics, such as packet delivery rates, effective transmission range, and neighborhood size (see Fig.~\ref{fig:neighborhood}).

\subsection{Vehicular Channel Emulations}

Performing experiments with real hardware in realistic environments is inherently the most realistic approach to characterize wireless vehicular channels. However, the cost and repeatability issues make this approach unfeasible for large-scale evaluations (e.g., involving tens or hundreds of vehicles). At the other end of the spectrum is channel simulation, which can ensure high repeatability, configurability, and manageability. 
However, designing a realistic channel simulator is a challenge, since the simulation environment needs to either be highly detailed to account for all aspects of the real system or it needs to make certain assumptions about the real world. 
Between the channel experimentation and simulation lies the channel emulation, where parts of the real communication systems are used in conjunction with the simulated ones, with the main goal of maintaining the repeatability and configurability of simulation environments, combined with a high level of realism of a testbed. One example is the Carnegie Mellon University (CMU) Wireless Emulator~\cite{judd2004repeatable}, where the emulator takes as input the signals generated by real devices, subjects them to simulated realistic signal propagation models and feeds the signals back into the real devices. The emulator was shown to behave realistically by comparing its output to the real-world measurements. In terms of the vehicular channel emulation, a detailed model for small-scale statistics of vehicular channels has been developed and implemented in NS-3 by Mittag et al.~\cite{mittag2011enabling}. The authors compared the results from the simulator with those generated by the CMU Wireless Emulator\cite{judd2004repeatable} and found a good match in terms of frame reception rate results. Therefore, provided that the size of the system is limited (up to a few dozen vehicles), the channel emulation is a feasible approach for a realistic and reproducible vehicular channel evaluation.

\subsection{Open Research Issues}
\subsubsection{Channel models for different vehicle types}
Vehicular channel measurements and modeling have primarily focused on personal cars (e.g.,~\cite{mangel11_2,boban14TVT,abbas12}). Studies dealing with  other types of vehicles (e.g., commercial vans, trucks, scooters, and public transportation vehicles) are rare, despite their considerably different dimensions and road dynamics. For example, the mobility of scooters and motorcycles is notably different from that of personal cars~\cite{shih11}. Combined with their smaller dimensions and lack of roof for antenna placement, the mobility of scooters indicates that the propagation characteristics for scooters can be significantly different from that of personal cars. Similarly, recent studies have shown that, in the same environment, commercial vans and trucks experience different channel propagation characteristics from the personal cars. This resulted in different reliable communication ranges and packet error rates~\cite{boban13}. Therefore, further studies are needed to investigate channel characteristics for vehicles other than personal cars.

\subsubsection{Under-explored environments}
While vehicular communications can take place in any scenario, signal propagation measurements are usually performed in the common environments (e.g., those in Fig.~\ref{fig:Environments}) and measurements in other environments, such as multi-level highways, tunnels, parking garages, bridges, and roundabouts are quite rare. For example, V2V signal propagation measurements in a parking garage has been performed in one study to date~\cite{sun2013parking}; similar for measurements in tunnels~\cite{maier2012channel,Fayziyev2014channel} and on-bridge environments~\cite{bernardo13}. Further measurements and modeling studies are particularly necessary for environments with distinct applications use-cases that can occur in them (e.g., service updates in parking garages, cooperative awareness functionalities without GPS in tunnels, etc.).

\subsubsection{Vehicle-to-X channels}
Despite significant differences between V2V, V2I, and V2P communications, the propagation characteristics of V2I and V2P channels are not as well-researched as V2V channels -- for example, all models described in Table~\ref{tab:Models} focus on V2V communications. In addition to V2I measurement campaigns~\cite{gozalvez12,tan08}, there are only a few dedicated V2I channel models. Acosta-Marum and Ingram~\cite{acosta07} developed a TDL model to capture the joint Doppler-delay characteristics of V2I channels. The models are based on extensive measurements for urban, suburban, and highway environments. Part of the reason is that V2I systems resemble existing cellular systems, where one of the communicating entity (base-station) is stationary while the other (user equipment) is mobile. However, typical positioning of static (infrastructure) nodes in V2I communications is unique for vehicular communications: on highways, road side units (RSUs) will be placed close to the road at heights considerably lower than that of cellular base stations (see, e.g., current efforts within the Amsterdam Group: \url{https://amsterdamgroup.mett.nl}). In urban areas, the most beneficial locations are near large intersections. Furthermore, a study performed by Gozalvez et al.~\cite{gozalvez12} showed that V2I communications in urban areas is highly variable, with both static and mobile objects creating a considerably changing channel over both space and time. Therefore, there exists a need for further studies to investigate V2I channels. 

In terms of V2P communications, recent studies by Wu et al.~\cite{wu2014cars} and Anaya et al.~\cite{anaya2014vehicle} have explored the basic channel properties of V2I links. Channel models for different communication technologies that can enable V2P communications (e.g., DSRC, WiFi, and cellular-based systems) need to be explored further. Therefore, there is much work to be done in order to fully understand and model the V2P communication channels.

\subsubsection{Vehicle-to-X and 5G}
As the recent research efforts on future 5G cellular networks start to look more deeply into ITS-related applications~\cite{osseiran2014scenarios}, it is reasonable to expect gradual convergence of the efforts on the channel modeling for Vehicle-to-X (V2X) and 5G systems. For example, as the delay requirement becomes more stringent for many 5G scenarios, the proposed system overcomes the main obstacle for use in a vehicular setting (i.e., lack of low-latency guarantees). Initial steps needed for enabling V2X systems through 5G, along with the related requirements for channel modeling, are discussed by Kyrolainen et al.~\cite{kyrolainen2014channel}. Additionally, when applied to highly mobile terminals, Device-to-Device (D2D) concept in 5G systems shares many similarities with V2X communications; therefore, efforts on modeling D2D channels (e.g.,~\cite{nurmela2013channel}) can benefit from the existing V2V channel modeling work and vice versa.

\section{Conclusions}
\label{sec:Conclusion}
This paper provides a survey of recent developments in the area of propagation and channel modeling for vehicular communications. We pay special attention to the usability aspects of the models, including their suitability for large scale evaluations of protocols and applications for future Cooperative Intelligent Transportation Systems. We first discuss the key channel characteristics that distinguish the vehicular communications from other types of wireless communications. Next, based on the distinguishing features, we classify and summarize the state of the art vehicular channel and propagation models based on the propagation mechanisms they model and their implementation approach. In addition, we provide guidelines for choosing a suitable channel model, depending on the type of protocol or application under investigation and taking into account the availability of geographical information and processing power available for simulation execution. 
Finally, we discuss the less-explored aspects of vehicular channel modeling and point out the areas where further research efforts are required.

\bibliographystyle{IEEEtran}

\begin{thebibliography}{10}
\providecommand{\url}[1]{#1}
\csname url@samestyle\endcsname
\providecommand{\newblock}{\relax}
\providecommand{\bibinfo}[2]{#2}
\providecommand{\BIBentrySTDinterwordspacing}{\spaceskip=0pt\relax}
\providecommand{\BIBentryALTinterwordstretchfactor}{4}
\providecommand{\BIBentryALTinterwordspacing}{\spaceskip=\fontdimen2\font plus
\BIBentryALTinterwordstretchfactor\fontdimen3\font minus
  \fontdimen4\font\relax}
\providecommand{\BIBforeignlanguage}[2]{{%
\expandafter\ifx\csname l@#1\endcsname\relax
\typeout{** WARNING: IEEEtran.bst: No hyphenation pattern has been}%
\typeout{** loaded for the language `#1'. Using the pattern for}%
\typeout{** the default language instead.}%
\else
\language=\csname l@#1\endcsname
\fi
#2}}
\providecommand{\BIBdecl}{\relax}
\BIBdecl

\bibitem{molisch2009survey}
A.~Molisch, F.~Tufvesson, J.~Karedal, and C.~Mecklenbr{\"a}uker, ``A survey on
  vehicle-to-vehicle propagation channels,'' \emph{IEEE Wireless
  Communications}, vol.~16, no.~6, pp. 12--22, 2009.

\bibitem{mecklenbrauker2011vehicular}
C.~F. Mecklenbrauker, A.~F. Molisch, J.~Karedal, F.~Tufvesson, A.~Paier,
  L.~Bernado, T.~Zemen, O.~Klemp, and N.~Czink, ``Vehicular channel
  characterization and its implications for wireless system design and
  performance,'' \emph{Proceedings of the IEEE}, vol.~99, no.~7, pp.
  1189--1212, 2011.

\bibitem{cheng09}
C.-X. Wang, X.~Cheng, and D.~I. Laurenson, ``Vehicle-to-vehicle channel
  modeling and measurements: Recent advances and future challenges,''
  \emph{IEEE Communications Magazine}, vol.~47, no.~11, pp. 96--103, November
  2009.

\bibitem{euPress}
\BIBentryALTinterwordspacing
``{New connected car standards put Europe into top gear},'' February 2014.
  [Online]. Available:
  \url{http://europa.eu/rapid/press-release_IP-14-141_en.htm}
\BIBentrySTDinterwordspacing

\bibitem{usdotNHTSA}
\BIBentryALTinterwordspacing
``{U.S. Department of Transportation Announces Decision to Move Forward with
  Vehicle-to-Vehicle Communication Technology for Light Vehicles},'' February
  2014. [Online]. Available:
  \url{http://www.nhtsa.gov/About+NHTSA/Press+Releases/2014/USDOT+to+Move+Forward+with+Vehicle-to-Vehicle+Communication+Technology+for+Light+Vehicles}
\BIBentrySTDinterwordspacing

\bibitem{abbas12}
T.~Abbas, K.~Sj{\"o}berg, J.~Karedal, and F.~Tufvesson, ``A measurement based
  shadow fading model for vehicle-to-vehicle network simulations,'' \emph{arXiv
  preprint arXiv:1203.3370v2}, 2012.

\bibitem{karedal09}
J.~Karedal, F.~Tufvesson, N.~Czink, A.~Paier, C.~Dumard, T.~Zemen,
  C.~Mecklenbr{\"a}uker, and A.~Molisch, ``A geometry-based stochastic {MIMO}
  model for vehicle-to-vehicle communications,'' \emph{IEEE Transactions on
  Wireless Communications}, vol.~8, no.~7, pp. 3646--3657, July 2009.

\bibitem{paschalidis11}
P.~Paschalidis, K.~Mahler, A.~Kortke, M.~Peter, and W.~Keusgen, ``Pathloss and
  multipath power decay of the wideband car-to-car channel at 5.7 {GHz},'' in
  \emph{IEEE Vehicular Technology Conference (VTC Spring)}, May 2011, pp. 1--5.

\bibitem{Cheng2007}
L.~Cheng, B.~E. Henty, D.~D. Stancil, F.~Bai, and P.~Mudalige, ``Mobile
  vehicle-to-vehicle narrow-band channel measurement and characterization of
  the 5.9 {GHz} {Dedicated Short Range Communication} {(DSRC)} frequency
  band,'' \emph{IEEE Journal on Selected Areas in Communications}, vol.~25,
  no.~8, pp. 1501--1516, Oct. 2007.

\bibitem{mangel11_2}
T.~Mangel, O.~Klemp, and H.~Hartenstein, ``A validated 5.9 {GHz}
  non-line-of-sight path-loss and fading model for inter-vehicle
  communication,'' in \emph{11th International Conference on ITS
  Telecommunications (ITST)}, August 2011, pp. 75 --80.

\bibitem{renaudin2008wideband}
O.~Renaudin, V.~Kolmonen, P.~Vainikainen, and C.~Oestges, ``Wideband {MIMO}
  car-to-car radio channel measurements at 5.3 {GHz},'' in \emph{IEEE 68th
  Vehicular Technology Conference (VTC Fall)}, 2008, pp. 1--5.

\bibitem{acosta07}
G.~Acosta-Marum and M.~Ingram, ``Six time- and frequency-selective empirical
  channel models for vehicular wireless {LANs},'' \emph{IEEE Vehicular
  Technology Magazine}, vol.~2, no.~4, pp. 4 --11, December 2007.

\bibitem{tan08}
I.~Tan, W.~Tang, K.~Laberteaux, and A.~Bahai, ``Measurement and analysis of
  wireless channel impairments in {DSRC} vehicular communications,'' in
  \emph{IEEE International Conference on Communications (ICC)}, May 2008, pp.
  4882--4888.

\bibitem{anaya2014vehicle}
J.~J. Anaya, P.~Merdrignac, O.~Shagdar, F.~Nashashibi, and J.~E. Naranjo,
  ``Vehicle to pedestrian communications for protection of vulnerable road
  users,'' in \emph{IEEE Intelligent Vehicles Symposium}, 2014, pp. 1037--1042.

\bibitem{rappaport96}
T.~S. Rappaport, \emph{Wireless Communications: Principles and Practice}.\hskip
  1em plus 0.5em minus 0.4em\relax Prentice Hall, 1996.

\bibitem{boban14TVT}
M.~Boban, J.~Barros, and O.~Tonguz, ``Geometry-based vehicle-to-vehicle channel
  modeling for large-scale simulation,'' \emph{IEEE Transactions on Vehicular
  Technology}, vol.~63, no.~9, pp. 4146--4164, Nov 2014.

\bibitem{vlastaras2014impact}
D.~Vlastaras, T.~Abbas, M.~Nilsson, R.~Whiton, M.~Olback, and F.~Tufvesson,
  ``Impact of a truck as an obstacle on vehicle-to-vehicle communications in
  rural and highway scenarios,'' in \emph{IEEE 6th International Symposium on
  Wireless Vehicular Communications (WiVeC)}, 2014, pp. 1--6.

\bibitem{maurer04_2}
J.~Maurer, T.~Fugen, T.~Schafer, and W.~Wiesbeck, ``{A new inter-vehicle
  communications (IVC) channel model},'' in \emph{IEEE Vehicular Technology
  Conference (VTC-Fall)}, September 2004, pp. 9--13.

\bibitem{sun2013parking}
R.~Sun, D.~W. Matolak, and P.~Liu, ``Parking garage channel characteristics at
  5 {GHz} for {V2V} applications,'' in \emph{IEEE 78th Vehicular Technology
  Conference (VTC Fall)}, 2013, pp. 1--5.

\bibitem{00parsons}
J.~D. Parsons, \emph{The Mobile Radio Propagation Channel}.\hskip 1em plus
  0.5em minus 0.4em\relax John Wiley \& Sons, 2000.

\bibitem{sen08}
I.~Sen and D.~W. Matolak, ``Vehicle-vehicle channel models for the 5-{GHz}
  band,'' \emph{IEEE Transactions on Intelligent Transportation Systems},
  vol.~9, no.~2, pp. 235--245, June 2008.

\bibitem{pilosu2011radii}
L.~Pilosu, F.~Fileppo, and R.~Scopigno, ``{RADII}: a computationally affordable
  method to summarize urban ray-tracing data for {VANETs},'' in
  \emph{International Conference on Wireless Communications, Networking and
  Mobile Computing (WiCOM)}, 2011, pp. 1--6.

\bibitem{boban11}
M.~Boban, T.~T.~V. Vinhoza, M.~Ferreira, J.~Barros, and O.~K. Tonguz, ``Impact
  of vehicles as obstacles in vehicular ad hoc networks,'' \emph{IEEE Journal
  on Selected Areas in Communications}, vol.~29, no.~1, pp. 15--28, January
  2011.

\bibitem{wang12}
X.~Wang, E.~Anderson, P.~Steenkiste, and F.~Bai, ``Improving the accuracy of
  environment-specific vehicular channel modeling,'' in \emph{Proceedings of
  the seventh ACM international workshop on Wireless network testbeds,
  experimental evaluation and characterization, {WiNTECH '12}}.\hskip 1em plus
  0.5em minus 0.4em\relax New York, NY, USA: ACM, 2012, pp. 43--50.

\bibitem{bernardo13}
L.~Bernad{\'o}, T.~Zemen, F.~Tufvesson, A.~F. Molisch, and C.~F.
  Mecklenbr{\"a}uker, ``Delay and {Doppler} spreads of non-stationary vehicular
  channels for safety relevant scenarios,'' \emph{CoRR}, vol. abs/1305.3376,
  2013.

\bibitem{dhoutaut06}
D.~Dhoutaut, A.~Regis, and F.~Spies, ``Impact of radio propagation models in
  vehicular ad hoc networks simulations,'' \emph{VANET 06: Proceedings of the
  3rd international workshop on Vehicular ad hoc networks}, pp. 69--78, 2006.

\bibitem{openstreetmap}
M.~Haklay and P.~Weber, ``{OpenStreetMap}: User-generated street maps,''
  \emph{IEEE Pervasive Computing}, vol.~7, no.~4, pp. 12--18, 2008.

\bibitem{judd2004repeatable}
G.~Judd and P.~Steenkiste, ``Repeatable and realistic wireless experimentation
  through physical emulation,'' \emph{ACM SIGCOMM Computer Communication
  Review}, vol.~34, no.~1, pp. 63--68, 2004.

\bibitem{mittag2011enabling}
J.~Mittag, S.~Papanastasiou, H.~Hartenstein, and E.~G. Str{\"o}m, ``Enabling
  accurate cross-layer {PHY/MAC/NET} simulation studies of vehicular
  communication networks,'' \emph{Proceedings of the IEEE}, vol.~99, no.~7, pp.
  1311--1326, 2011.

\bibitem{shih11}
O.~Shih, H.~Tsai, H.~Lin, and A.~Pang, ``A rule-based mixed mobility model for
  cars and scooters (poster),'' in \emph{IEEE Vehicular Networking Conference
  (VNC)}, 2011, pp. 198--205.

\bibitem{boban13}
M.~Boban, R.~Meireles, J.~Barros, P.~A. Steenkiste, and O.~K. Tonguz, ``{TVR} -
  tall vehicle relaying in vehicular networks,'' \emph{IEEE Transactions on
  Mobile Computing}, vol.~13, no.~5, pp. 1118--1131, May 2014.

\bibitem{maier2012channel}
G.~Maier, A.~Paier, and C.~Mecklenbr{\"a}uker, ``Channel tracking for a
  multi-antenna {ITS} system based on vehicle-to-vehicle tunnel measurements,''
  in \emph{19th IEEE Symposium on Communications and Vehicular Technology in
  the Benelux (SCVT)}, 2012, pp. 1--6.

\bibitem{Fayziyev2014channel}
A.~Fayziyev, M.~PŠtzold, E.~Masson, Y.~Cocheril, and M.~Berbineau, ``A
  measurement-based channel model for vehicular communications in tunnels,'' in
  \emph{IEEE Wireless Communications and Network Conference (WCNC)}, 2014, pp.
  128--133.

\bibitem{gozalvez12}
J.~Gozalvez, M.~Sepulcre, and R.~Bauza, ``{IEEE} 802.11p vehicle to
  infrastructure communications in urban environments,'' \emph{IEEE
  Communications Magazine}, vol.~50, no.~5, pp. 176 --183, May 2012.

\bibitem{wu2014cars}
X.~Wu, R.~Miucic, S.~Yang, S.~Al-Stouhi, J.~Misener, S.~Bai, and W.-h. Chan,
  ``Cars talk to phones: A dsrc based vehicle-pedestrian safety system,'' in
  \emph{IEEE Vehicular Technology Conference (VTC Fall)}.\hskip 1em plus 0.5em
  minus 0.4em\relax IEEE, 2014, pp. 1--7.

\bibitem{osseiran2014scenarios}
A.~Osseiran, F.~Boccardi, V.~Braun, K.~Kusume, P.~Marsch, M.~Maternia,
  O.~Queseth, M.~Schellmann, H.~Schotten, H.~Taoka \emph{et~al.}, ``Scenarios
  for 5g mobile and wireless communications: the vision of the metis project,''
  \emph{IEEE Communications Magazine}, vol.~52, no.~5, pp. 26--35, 2014.

\bibitem{kyrolainen2014channel}
J.~Kyr{\"o}l{\"a}inen, P.~Ky{\"o}sti, J.~Meinil{\"a}, V.~Nurmela,
  L.~Raschkowski, A.~Roivainen, and J.~Ylitalo, ``Channel modelling for the
  fifth generation mobile communications,'' in \emph{Proc. 8th European
  Conference on Antennas and Propagation, EuCAP}, 2014.

\bibitem{nurmela2013channel}
V.~Nurmela, T.~J{\"a}ms{\"a}, P.~Ky{\"o}sti, V.~Hovinen, and J.~Medbo,
  ``Channel modelling for device-to-device scenarios,'' \emph{COST IC}, vol.
  1004, pp. 1--6, 2013.

\end{thebibliography}

\section{Biographies}

{\bf Wantanee Viriyasitavat} is a lecturer in the Faculty of Information and Communication Technology at Mahidol University, Bangkok, Thailand. During 2012-2013, she was a Research Scientist in the Department of Electrical and Computer Engineering at Carnegie Mellon University (CMU), Pittsburgh, PA. She received her B.S./M.S., and Ph.D. degrees in electrical and computer engineering from CMU in 2006 and 2012, respectively. Between 2007-2012, she was a
Research Assistant at Carnegie Mellon University, where she was a member of General Motors Collaborative Research Laboratory (CRL) and was working on the design of a routing framework for safety and non-safety applications of vehicular ad hoc wireless networks (VANETs). Her research interests include traffic mobility modeling,
network connectivity analysis, and protocol design for wireless ad hoc networks.

{\bf Mate Boban} is a Research Scientist at NEC Laboratories Europe. He holds a Ph.D. in Electrical and Computer Engineering from Carnegie Mellon University and a Diploma in Informatics from University of Zagreb. He is an alumnus of the Fulbright Scholar Program. His current research is in the areas of Intelligent Transportation Systems, wireless communications, and networking. He received the Best Paper Award at IEEE VTC 2014-Spring and at IEEE VNC 2014. More information can be found on his website http://mateboban.net.

{\bf Hsin-Mu Tsai} is an assistant professor in Department of Computer Science and Information Engineering and Graduate Institute of Networking and Multimedia at National Taiwan University. He received his B.S.E in Computer Science and Information Engineering from National Taiwan University in 2002, and his M.S. and Ph.D. in Electrical and Computer Engineering from Carnegie Mellon University in 2006 and 2010, respectively. Dr. Tsai's recognitions include 2014 Intel Labs Distinguished Collaborative Research Award, 2013 Intel Early Career Faculty Award (the first to receive this honor outside of North America and Europe), and National Taiwan University's Distinguished Teaching Award. Dr. Tsai served as the workshop co-chair for the first ACM Visible Light Communication System (VLCS) Workshop in 2014, and TPC co-chair for ACM VANET 2013. His research interests include vehicular networking and communications, wireless channel and link measurements, vehicle safety systems, and visible light communications.

{\bf Athanasios V.Vasilakos} (M'00-SM'11) is currently Professor with Lulea University of Technology, Sweden. He has authored or co-authored over 200 technical papers in major international journals and conferences. He is author/coauthor of five books and 20 book chapters in the areas of communications.
He served or is serving as an Editor or/and Guest Editor for many technical journals, such as the IEEE Transactions on Network and Services Management, IEEE Transactions on Cloud Computing, IEEE Transactions on Information Forensics and Security, IEEE Transactions on Cybernetics, IEEE Transactions on Information Technology in Biomedicine, ACM Transactions on Autonomous and Adaptive Systems, IEEE Journal on Selected Areas in Communications. He is also General Chair of the European Alliances for Innovation (www.eai.eu).

\end{document}